# Dynamical Localization and Absolute Negative Conductance in a AC Driven Double Quantum Well.

Ramón Aguado and Gloria Platero

*Instituto de Ciencia de Materiales (CSIC), Cantoblanco, 28049 Madrid, Spain.*

## Abstract

We analyze the I-V characteristic of a double quantum well integrated in an antenna and driven by THz radiation. We propose a microscopic model to calculate the sequential current including selfconsistently the Coulomb interaction in a mean field approximation. Additional features in the current appear which depend not only on the frequency and intensity of the ac field but also on the charge in the wells. We observe, for some intensities and frequencies of the ac field, dynamical localization, i.e., quenching of the transmission, and, in the linear response regime, absolute negative conductance. We compare our calculations with recent experiments in driven double wells.

73.20.Dx,73.40.Gk

The effect of an ac field on the transport properties of nanostructures has been a subject of increasing interest in the last years. The experiments on tunneling through a metal-superconductor diode [1] irradiated with microwaves indicated that the external field assists the tunneling openning new effective channels. Recently transport experiments have been performed in semiconductor nanostructures in different ac fields: The photoassisted current through a double barrier (DB) irradiated with FIR light shows [2,3] additional features which can be explained in terms of the coupling of different electronic states induced by the electromagnetic field [4]. The effect of an AC potential on the conductance and current in quantum dots (QD's) [5–7] , superlattices (SL's) [8] and double quantum wells (DQW's) [9] has been measured and new physical features as dynamical localization and absolute negative conductance have been observed [8]. Also Rabi transitions between discrete states and electron pumping in double quantum dots (DQD's) has been analyzed within the Keldysh formalism [10]. In this work we propose a model for analyzing the time-average current in a triple barrier integrated in an antenna and driven by THz radiation. In this configuration, the irradiated antenna produces an oscillatory signal between the left and right lead, i.e., a time dependent bias drops between the emitter and collector. It is important to stress that this configuration of the ac field is different than the one corresponding to homogeneous irradiation of the sample [2–4]. In order to account with the scattering due to roughness at the interfaces, phonons or impurities we analyze sequential tunnel: we calculate the current from the emitter to the left well $J_{e,1}$, from this one to the right well $J_{1,2}$ and from the right well to the collector $J_{2,c}$. Current conservation determinates the Fermi energies in the wells and the sequential current. The effect of electron-electron interaction in these quasi-3D structures can be treated as a perturbation to the non-interacting system within a mean field approximation [11,12]. The charge accumulated in the wells produces an additional electrostatic potential which modifies their electronic structure, transmission coefficient and current. We solve selfconsistently the Schrodinger equation including the ac field and the Poisson equation. The induced electrostatic potential does not drop linearly and accumulation and depletion layers in the emitter and collector are built up. Details of the electrostatic model are given elsewhere [12].

The hamiltonian of the DQW driven by an ac potential is:

$$H(t) = \sum_{k_i \epsilon L,R} E_{k_i}(t) c^\dagger_{\mathbf{k_i}} c_{\mathbf{k_i}} + \sum_{k_j \epsilon 1,2} E_{k_j}(t) d^\dagger_{\mathbf{k_j}} d_{\mathbf{k_j}}$$
$$+ \sum_{k_i k_j \begin{cases} i=L & j=1 \\ i=R & j=2 \end{cases}} (T_{k_i k_j} c^\dagger_{\mathbf{k_i}} d_{\mathbf{k_j}} + H.c) + \sum_{k_1 k_2}(T_{k_1 k_2} d^\dagger_{\mathbf{k_1}} d_{\mathbf{k_2}} + H.c) \quad (1)$$

Here $T_{k_i k_j}, T_{k_1 k_2}$ are the tunneling matrix elements (leads-well and well-well respectively) [13]. The ac field induces a time dependent bias between the emitter and collector producing a dipole around the central region. The electronic energies become time dependent $E_{k_i}(t) = E^0_{k_i} + eFz_i cos\omega_0 t$, where $z_i$ is taken constant (its mean position) in each spatial region. It implies that the mixing of electronic states within each spatial region due to the position operator $\hat{\mathbf{Z}}$ is neglected. This approximation, which was assumed in the model of Tien and Gordon [14] is reasonable for this configuration of the ac field, where the time dependent potential drops just between the leads and of course in non resonant conditions, i.e, for $\omega_0 \neq \frac{\Delta}{\hbar}$ ($\Delta$ being the subband spacing). The operators acquire phase factors of the form $\mathbf{c_{ki}}(t) = e^{-\frac{i}{\hbar}E_{ki}t}e^{-\frac{ieFz_i}{\hbar\omega_0}sin\omega_0 t}\mathbf{c_{ki}} = e^{-\frac{i}{\hbar}E_{ki}t}\sum_{n=-\infty}^{\infty}J_n(\frac{eFz_i}{\hbar\omega_0})e^{-in\omega_0 t}\mathbf{c_{ki}}$.

In the case of sequential tunneling first order in perturbation theory gives a good description of the tunneling (the coherent tunneling, i.e, infinite order in the tunneling perturbation has been solved in [13]). We obtain for the transition probabilities:

$$P_1(E_e, E_1) = \frac{2\pi}{\hbar}\sum_{n=-\infty}^{\infty}J_n^2(\beta_{e1})(\frac{2\hbar^4}{m^{*2}})\frac{k_e}{L}[\frac{k_e k_1^2(\alpha_1 + \alpha_1')^2 e^{-2\alpha_1 d_1}}{(d_2 + \frac{1}{\alpha_1'} + \frac{1}{\alpha_2})(k_e^2 + \alpha_1^2)(k_1^2 + \alpha_1'^2)}]$$

$$\frac{2\pi^2}{S^2}\delta\vec{k}_{\parallel}\delta(E_e - E_1 + n\hbar\omega_0)$$

$$P_2(E_1, E_2) = \frac{2\pi}{\hbar}\sum_{m=-\infty}^{\infty}J_m^2(\beta_{12})(\frac{\hbar^4}{m^{*2}})[\frac{k_1^2 k_2^2(\alpha_2 + \alpha_2')^2 e^{-2\alpha_2 d_3}}{(d_2 + \frac{1}{\alpha_1'} + \frac{1}{\alpha_2})(d_4 + \frac{1}{\alpha_2'} + \frac{1}{\alpha_3})(k_1^2 + \alpha_2^2)(k_2^2 + \alpha_2'^2)}]$$

$$\frac{2\pi^2}{S^2}\delta\vec{k}_{\parallel}\delta(E_1 - E_2 + m\hbar\omega_0)$$

$$P_3(E_2, E_c) = \frac{2\pi}{\hbar}\sum_{p=-\infty}^{\infty}J_n^2(\beta_{2c})(\frac{2\hbar^4}{m^{*2}})\frac{k_c}{L}[\frac{k_2^2 k_c(\alpha_3 + \alpha_3')^2 e^{-2\alpha_3 d_5}}{(d_4 + \frac{1}{\alpha_2'} + \frac{1}{\alpha_3})(k_2^2 + \alpha_3^2)(k_c^2 + \alpha_3'^2)}]$$

$$\frac{2\pi^2}{S^2}\delta\vec{k}_{\parallel}\delta(E_2 - E_c + p\hbar\omega_0) \qquad (2)$$

Where $E_k = \frac{\hbar^2 k_\parallel^2}{2m^*} + \epsilon_{k_z}$, the rest of parameters are shown in fig.1, $\beta_{e1} = \frac{eF(z_1 - z_e)}{\hbar\omega_0}$, $\beta_{12} = \frac{eF(z_2 - z_1)}{\hbar\omega_0}$ and $\beta_{2c} = \frac{eF(z_c - z_2)}{\hbar\omega_0}$ are related with the ac potential drops between the different regions, and the terms in brackets are the inelastic transmissions $T_1(\epsilon_e, \epsilon_1), T_2(\epsilon_1, \epsilon_2), T_3(\epsilon_2, \epsilon_c)$ through each barrier. The time describing the relaxation between the states due to coupling with phonons is $\sim 10^{-13}s$ [15]. The DOS in the growth direction is described then as Lorentzians of 1 meV half width. In formula (2) irreversibility is assumed. In this case the Fermi Golden Rule is valid [16]: it is clearly the case for $P_1$ and $P_3$ where there is a continuum of states involved. In the case of the transmission probability from the left to right well we assume irreversibility due to the broadening of the DOS induced by scattering as well as to the 2D continuous in-plane DOS. On top on that, if the tunneling time associated to the outer barrier is shorter than the one corresponding to the barrier separating the two wells the process of tunnel becomes irreversible. However, in the case of an isolated DQD, non connected with a continuous DOS at the leads and in the ideal case were no scattering occurs the tunnel is a reversible process and a description in terms of Rabi transitions should be considered [10,16]. The expression for the currents is:

$$J_{e1} = \frac{2ek_BT}{\pi^2\hbar} \sum_{n=-\infty}^{\infty} J_n^2(\beta_{e1}) \int \frac{\gamma}{[(\epsilon_e + n\hbar\omega_0 - \epsilon_{r1})^2 + \gamma^2]}$$

$$\times T_1(\epsilon_e, \epsilon_e + n\hbar\omega_0) ln[\frac{1 + e^{\frac{(\epsilon_F - \epsilon_e)}{k_BT}}}{1 + e^{\frac{(\epsilon_{w1} - \epsilon_e - n\hbar\omega_0)}{k_BT}}}]d\epsilon_e$$

$$J_{12} = \frac{2e\hbar k_BT}{\pi^2 m^*} \sum_{m=-\infty}^{\infty} J_m^2(\beta_{12}) \int \frac{\gamma}{[(\epsilon_1 - \epsilon_{r1})^2 + \gamma^2]} \frac{\gamma}{[(\epsilon_1 - \epsilon_{r2} + m\hbar\omega_0)^2 + \gamma^2]}$$

$$\times T_2(\epsilon_1, \epsilon_1 + m\hbar\omega_0) ln[\frac{1 + e^{\frac{(\epsilon_{w1} - \epsilon_1)}{k_BT}}}{1 + e^{\frac{(\epsilon_{w2} - \epsilon_1 - m\hbar\omega_0)}{k_BT}}}]d\epsilon_1$$

$$J_{2c} = \frac{2ek_BT}{\pi^2\hbar} \sum_{p=-\infty}^{\infty} J_p^2(\beta_{2c}) \int \frac{\gamma}{[(\epsilon_2 - \epsilon_{r2})^2 + \gamma^2]}$$

$$\times T_3(\epsilon_2, \epsilon_2 + p\hbar\omega_0) ln[\frac{1 + e^{\frac{(\epsilon_{w2} - \epsilon_2)}{k_BT}}}{1 + e^{\frac{(\epsilon_F - eV - \epsilon_2 - p\hbar\omega_0)}{k_BT}}}]d\epsilon_2 \quad (3)$$

Where $\epsilon_{r1}$ and $\epsilon_{r2}$ are the ground states of the left and right wells. For simplicity we restrict the previous equations to the tunneling between ground to ground state (the generalization to excited well states is straightforward). The Fermi energies $\epsilon_{w1}$ and $\epsilon_{w2}$ of the wells are obtained through the set of transcendental equations $J - J_{e1} = 0; J - J_{12} = 0; J - J_{2c} = 0$. At this point we solve the Poisson and Schrodinger equation selfconsistently to obtain the current. In order to compare with recent experiments [9], we analyze a sample consisting in two GaAs quantum wells 180 and $100\mathring{A}$ wide separated by a barrier of $Ga_{.7}Al_{.3}As$ of $60\mathring{A}$. The outer barriers are $35\mathring{A}$ thick (sample a). The emitter and collector are $n^+$ doped ($10^{18}cm^{-3}$). In figure (2.a) the J/V curve is shown with and without Coulomb interaction. We observe a main peak coming from the alignement of the two ground states of the two wells. Two additional peaks show up at both sides of the main peak symmetric in bias but asymmetric in intensity and correspond to induced one photon absorption and emission. The asymmetry in intensity takes place due to the increase in the transmission probability at higher bias and to the asymmetry of the charge accumulated in the wells (see fig. 2.b). The behaviour of J for different F and fixed $\omega_0$ is shown in fig. 3.a: as F increases new satellites show up indicating that multiphoton processes increase in probability, the main peak is reduced and the satellites become more intense. Also J is quenched for some frequencies and intensities of the field. This effect has been discussed [13,17,18] in terms of zeros of the $J_0$ Bessel functions: as the argument approaches the first zero ( $\sim 2.4$) dynamical localization of electrons takes place, independently of the transparence of the barriers. The experiments [9] are very similar to our results. However we did not observe the small shift that the peaks experience as the intensity increases and which is interpretated as heating of the contact regions [9]. As the power increases, the effect of the ac field on the charge accumulated in the wells increases too. It would be expected [11] that for high intensities the ac field would affect the charge and the selfconsistent current explaining also the shift of the peaks. We do not observe an appreciable shift for intensities of the order of $10^5 \frac{V}{m}$ but it is observed for intensities at least one order of magnitude higher (as in the DB case [11]). There is also another effect which could explain this shift considering the dipole term as a quantum operator [4,19]. The term $eF\hat{Z}cos(\omega_0 t)$ will produce a shift of the energies: $\delta = \frac{e^2 F^2 \langle k_i|\hat{Z}|k_i\rangle^2}{\hbar \omega_0}$. Due to the smallness of the diagonal matrix elements of $\hat{Z}$ the shift becomes negligible [4]. J/V for different frequencies is shown in fig. 3.b. As $\omega_0$ increases the satellite peaks move far appart linearly from the main one which increases in intensity as the satellite peaks decrease. Finally we observe absolute negative conductance in a different sample: a $Ga_{.7}Al_{.3}As$ triple barrier of $50\mathring{A}$ and a DQW of GaAs of $150\mathring{A}$ and $n^+ = 10^{17} cm^{-3}$ (sample b). The reason for that is the asymmetry induced by the harmonic term $Fzcos\omega_0 t$ in the hamiltonian. This effect which has been discussed for different systems [17,20] has been observed in SL's [8] in the low bias region. The negative conductance induced by the ac field can be explained in terms of the Fermi distributions. Equation (3) shows that the supply function contains a superposition of effective dc voltages $\frac{\pm p\hbar\omega_0}{e}$ each one weighted nonlinearly by the Bessel functions. This effect is responsible of the negative conductance because the Fermi factors can be expressed in terms of effective Fermi energies (for instance, $E_{F_{effec}} = E_F - eV \pm p\hbar\omega_0$, in the collector). For some ranges of $F$ and $\omega_0$ the electrons are able to overcome the static bias ($eV \leq p_{max}\hbar\omega_0$) and electronic pumping at zero bias occurs [21]. We study the dependence of the absolute negative conductance with F including also the effect of Coulomb interaction (see fig. 4a,b). We observe that the charge interaction changes in some cases the sign of the differential conductance and the current. Therefore it modifies qualitatively the current in the low bias regime and has to be included to describe it properly. Also we observe that tunning F the differential conductance changes sign as well (fig. 4a) as the experimental evidence shows. Also in agreement with the experiments the conductance becomes negative at low bias increasing F and again positive as F increases further [8].

In conclusion, the ac sequential current through a triple barrier is evaluated including the Coulomb interaction, scattering effects and temperature. By means of our model we can explain the behaviour of the current versus voltage as a function of the ac frequency and intensity in good agreement with the experimental information. We analyze different samples and we observe interesting features in the current as dynamical localization and absolute negative conductance. Both effects can be controlled by tailoring the sample and harmonic potential configuration.

We acknowledge Dr. J. Iñarrea for enlighting discussions. One of us (R.A) acknowledge Fundación Universidad Carlos III de Madrid for financial support. This work has been supported by the CICYT (Spain) under contract MAT 94-0982-c02-02.

FIGURES

FIG. 1. Schematic diagram of sequential tunneling through a triple barrier. in an ac potential.

FIG. 2. a) J/V for ac sequential tunneling through a triple barrier (with and without selfconsistency)(inset without ac); b) 2D selfconsistent charge in the left and right well as a function of V (the inset shows the Fermi levels of the two wells ) for sample a.($F = 1.5 10^5 \frac{V}{m}, \omega_0 = 1.5 THz, T = 100^o K$).

FIG. 3. a) Selfconsistent J/V for an ac $\omega_0$=1.5 THz and different F ; b) at $F = 1.5 10^5 \frac{V}{m}$ and different ac frequencies (sample a) ($T = 100^o K$).

FIG. 4. a) Selfconsistent J/V for an ac $\omega_0$=1.5 THz and different F in the low bias regime (sample b); b) Same as a) but without Coulomb interaction ($T = 100^o K$).